\documentclass[letter, twocolumn]{jpsj3} 
\usepackage{txfonts}
\usepackage{epstopdf}

\title{Fermi surface and mass enhancement in KFe$_2$As$_2$ from de Haas-van Alphen effect measurements}

\author{Taichi \textsc{Terashima}$^{1, 5}$, Motoi \textsc{Kimata}$^{1, 5}$, Nobuyuki \textsc{Kurita}$^{1, 5}$, Hidetaka \textsc{Satsukawa}$^{1}$, Atsushi \textsc{Harada}$^{1}$, Kaori \textsc{Hazama}$^{1}$, Motoharu \textsc{Imai}$^{1, 5}$, Akira \textsc{Sato}$^{1}$, Kunihiro \textsc{Kihou}$^{2, 5}$, Chul-Ho \textsc{Lee}$^{2, 5}$, Hijiri \textsc{Kito}$^{2, 5}$, Hiroshi \textsc{Eisaki}$^{2, 5}$, Akira \textsc{Iyo}$^{2, 5}$, Taku \textsc{Saito}$^{3}$, Hideto \textsc{Fukazawa}$^{3, 5}$, Yoh \textsc{Kohori}$^{3, 5}$, Hisatomo \textsc{Harima}$^{4, 5}$, and Shinya \textsc{Uji}$^{1, 5}$}

\inst{$^{1}$National Institute for Materials Science, Tsukuba, Ibaraki 305-0003, Japan\\
$^{2}$National Institute of Advanced Industrial Science and Technology (AIST), Tsukuba, Ibaraki 305-8568, Japan\\
$^{3}$Department of Physics, Chiba University, Chiba 263-8522, Japan\\
$^{4}$Department of Physics, Graduate School of Science, Kobe University, Kobe, Hyogo 657-8501, Japan\\
$^{5}$JST, Transformative Research-Project on Iron Pnictides (TRIP), Chiyoda, Tokyo 102-0075, Japan}

\abst{We report on a band structure calculation and de Haas-van Alphen measurements of KFe$_2$As$_2$.  Three cylindrical Fermi surfaces are found.  Effective masses of electrons range from 6 to 18$m_e$, $m_e$ being the free electron mass.  Remarkable discrepancies between the calculated and observed Fermi surface areas and the large mass enhancement ($\gtrsim 3$) highlight the importance of electronic correlations in determining the electronic structures of iron pnicitide superconductors.}

\kword{iron pnictide superconductors, Fermi surface, de Haas-van Alphen effect}

\begin{document}
\maketitle

\newcommand{\ud}{\mathrm{d}}
\def\degree{\r{}}

The discovery of superconductivity at $T_c$ = 26 K in LaFeAs(O,~F) \cite{Kamihara08JACS} has given rise to intense experimental and theoretical efforts to elucidate the superconducting pairing mechanism and symmetry in iron pnictide superconductors (see Ref.~\citen{Ishida09JPSJ_review} for a recent review).  Since the development of realistic theories of the mechanism requires detailed knowledge of the Fermi surface (FS), experimental determination of the FS is highly desirable.

Accordingly, many angle-resolved photoemission spectroscopy (ARPES) studies have been performed\cite{Ishida09JPSJ_review}.  Their results show some level of agreement in the FS and band dispersion with conventional band structure calculations and moderate mass renormalization due to many-body effects.  On the other hand, measurements of de Haas-van Alphen (dHvA) or other quantum oscillations, which are bulk probes and allow accurate determination of the FS cross sections and effective masses $m^*$,  are rather limited.  dHvA measurements performed on the FeP compounds LaFePO \cite{Sugawara08JPSJ, Coldea08PRL} and SrFe$_2$P$_2$ \cite{Analytis09PRL} have shown that band shifts of up to $\sim$0.1 eV are necessary to bring band structure calculations into agreement with experiments and that the enhancement of effective masses over band ones is about two.  Since high $T_c$'s are found only in FeAs compounds, dHvA studies of FeAs compounds are more desired.  However, because of the structural/magnetic phase transitions, measurements on the alkaline-earth 122 parent compounds $A$Fe$_2$As$_2$ ($A$ = Ca, Sr, and Ba)\cite{Sebastian08JPCM, Analytis09PRB, Harrison09JPCM} have observed only small FS pockets, which makes it difficult to draw an overall picture of the electronic structures of these compounds.  Very recently, dHvA measurements have been performed on BaFe$_2$(As$_{1-x}$P$_x$)$_2$ for 0.41 $\leq x \leq$ 1 \cite{Shishido09condmat}.  As one goes from $x$ = 1 to 0.41, where $T_c \sim$ 25 K, the electron FS's shrink and the mass enhancement factor increases from $\sim$2 to $\sim$4.

KFe$_2$As$_2$ is an end member of the high-$T_c$ binary alloy (Ba$_{1-x}$K$_x$)Fe$_2$As$_2$ with the ThCr$_2$Si$_2$ structure and has $T_c \sim$3 K \cite{Rotter08ACIE, Sasmal08PRL}.  The low-temperature resistivity exhibits a clear $T^2$ dependence with a large coefficient of $A$ = 0.026 $\mu \Omega$cm/K$^2$ \cite{Terashima09JPSJKFA}, and specific heat measurements have found correspondingly large Sommerfeld coefficients: $\gamma_{exp}$ = 69 or 93 mJ/K$^2$mol-f.u. (f.u. = formula unit) for poly or single crystals, respectively \cite{Fukazawa09JPSJ_KFA, Fukazawa09PC}.  These indicate the existence of moderately large electron correlations.  $^{75}$As nuclear quadrupole resonance measurements have shown that spin fluctuations (SF's) are much suppressed (compared with the optimally doped compound) \cite{Fukazawa09JPSJ_KFA}.  The first ARPES measurement \cite{Sato09PRL} found $\alpha$ and $\beta$ hole FS's at the $\Gamma$ point in the Brillouin zone (BZ) and $\epsilon$ hole FS near the X point.  The $\alpha$ and $\beta$ FS's occupy 7 and 22\% of the BZ, while the $\epsilon$ FS occupies $4 \times 0.6$\% (because of the tetragonal symmetry, there are four $\epsilon$ cylinders in the BZ).  The authors assumed the double degeneracy of the $\beta$ FS.  The mass enhancement factor was found to be 2--4.  A more recent measurement \cite{Yoshida09PC}, which was performed on single crystals from the same source as our dHvA samples, has confirmed the basic FS topology but with slightly different FS sizes, and has revealed that the $\alpha$ FS, not $\beta$, is actually quasi-degenerate, i.e., two FS's with similar sizes exist.

We here present results of a band structure calculation and dHvA measurements of KFe$_2$As$_2$.  A substantial part of the large FS has been observed for the first time for the superconducting FeAs compounds.  Comparison with the band structure calculation will reveal the importance of electronic correlations in the FeAs compounds.

\begin{figure*}
\begin{center}
\includegraphics[width=17.5cm]{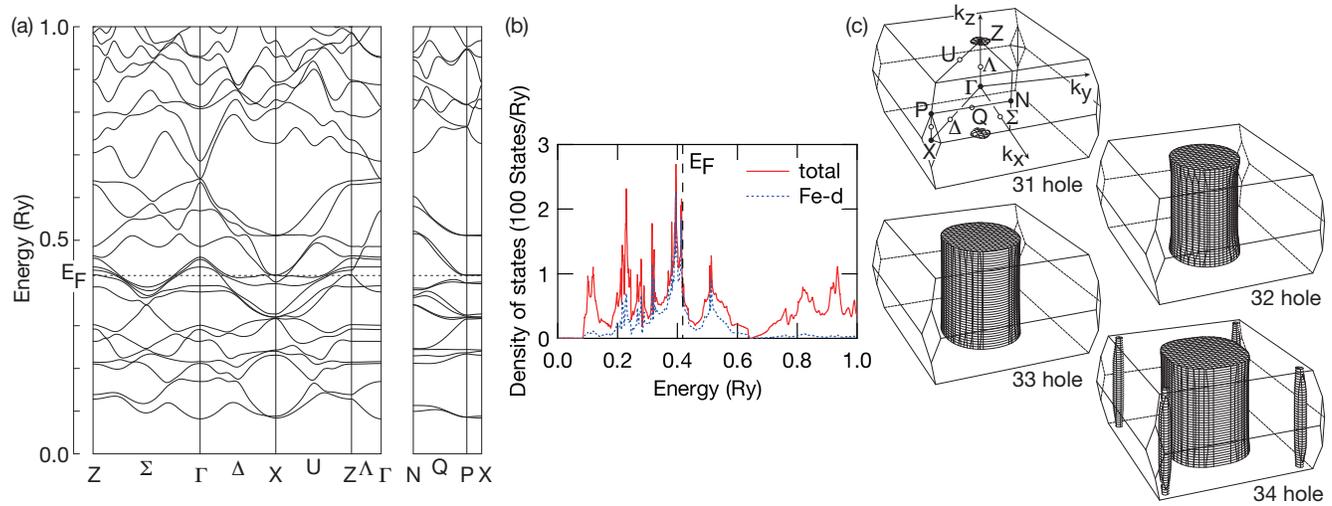}
\end{center}
\caption{\label{fig1}(color online).  (a) Calculated electronic band structure, (b) density of states, and (c) Fermi surface of KFe$_2$As$_2$.  Points of symmetry (solid circles) and lines of symmetry (open circles) in the Brillouin zone are explained in the top left figure of (c).}   
\end{figure*}

\begin{table}
\caption{\label{Tab1}Experimental and calculated Fermi surface parameters. $B\parallel c$.  $m_e$ is the free electron mass.}
\begin{tabular}{ccccccc}
\hline
 & \multicolumn{2}{c}{Experiment} & \multicolumn{3}{c}{Calculation} & \\
 \cline{2-3} \cline{4-6}
 & \multicolumn{1}{c}{$F$ (kT)} & \multicolumn{1}{c}{$\frac{m^*}{m_e}$} & Band & \multicolumn{1}{c}{$F$ (kT)} & \multicolumn{1}{c}{$\frac{m_{band}}{m_e}$} & \multicolumn{1}{c}{$\frac{m^*}{m_{band}}$} \\
\hline
\multicolumn{1}{c}{$\epsilon_l$} & 0.24 & 6.0(4) & \multicolumn{1}{c}{34 ($k_z$ = $\frac{2\pi}{c}$)} & 0.03 & 0.3 & 20 \\
\multicolumn{1}{c}{$\epsilon_h$} & 0.36 & 7.2(2) & \multicolumn{1}{c}{34 ($k_z$ = 0} & 0.10 & 0.3 & 24 \\
 &  &  & \multicolumn{1}{c}{31 (Z)} & 0.29 & 1.5 & \\
\multicolumn{1}{c}{$\alpha_l$} & 2.30 & 6.0(2) & \multicolumn{1}{c}{32 ($\Gamma$)} & 3.42 & 1.4 & 4.3 \\
\multicolumn{1}{c}{$\alpha_h$} & 2.39 & 6.5(2) & \multicolumn{1}{c}{32 (Z)} & 3.86 & 2.4 & 2.7 \\
\multicolumn{1}{c}{$\zeta_l$} & 2.89 & 8.5(2) & \multicolumn{1}{c}{33 ($\Gamma$)} & 4.67 & 2.2 & 3.9 \\
\multicolumn{1}{c}{$\zeta_h$} & 4.40 & 18(2) & \multicolumn{1}{c}{33 (Z)} & 4.88 & 2.6 & 6.9 \\
 &  &  & \multicolumn{1}{c}{34 ($\Gamma$)} & 5.82 & 2.6 & \\
 &  &  & \multicolumn{1}{c}{34 (Z)} & 6.03 & 2.9 & \\
\hline
\end{tabular}
\end{table}

\begin{figure}
\begin{center}
\includegraphics[width=8.5cm]{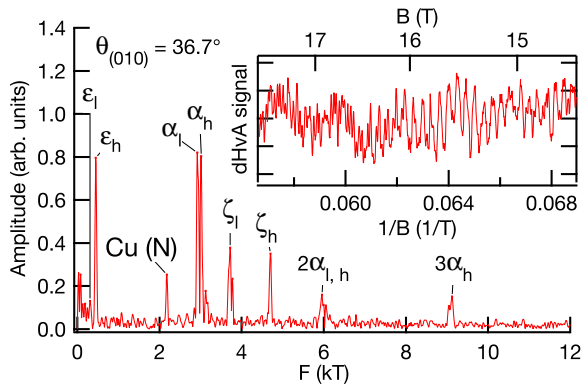}
\end{center}
\caption{\label{fig2}(color online).  Fourier transform of dHvA oscillations in inverse field at $\theta_{(010)}$ = 36.7\degree.  The used field range is between 10 and 17.65 T.  $T$ = 0.08 K.  dHvA frequencies are labeled with Greek letters.  The peak marked ``Cu(N)'' is due to the Cu neck oscillation coming from Cu wire of the pick-up coil.  A part of the raw oscillation data is shown in the inset.  The oscillation data at high fields near 17 T may appear noisier than low fields, but it is not due to noise.  It is because fast oscillations of the 3$\alpha_h$ frequency appear at high fields.}   
\end{figure}

\begin{figure*}
\begin{center}
\includegraphics[width=17.5cm]{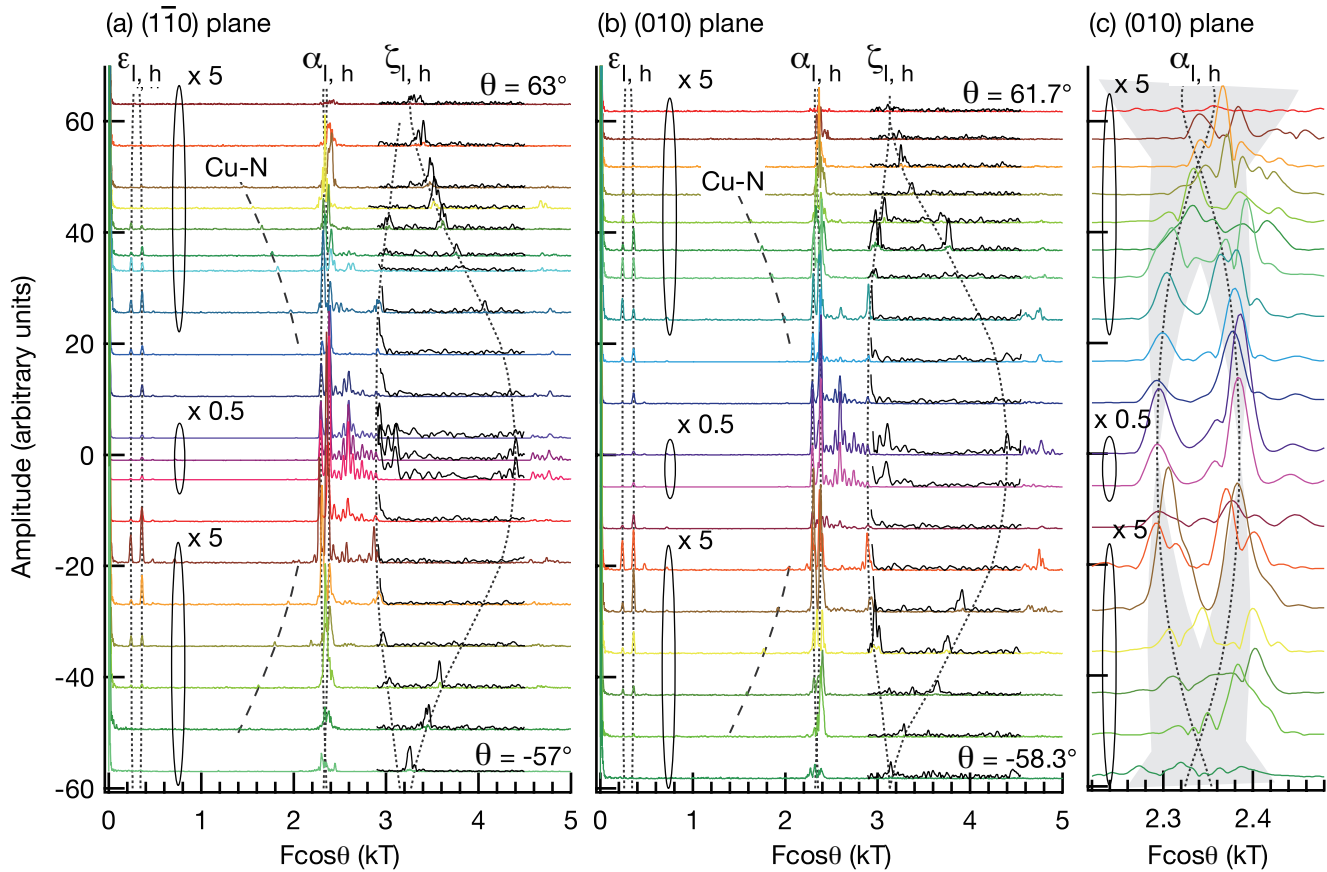}
\end{center}
\caption{\label{fig3}(color online).  Field-direction dependence of the Fourier transforms for the (a) $(1\overline10)$ and (b) (010) planes.  (c) is an enlarged view of (b) for a frequency range of $\alpha_{l, h}$.  Note that the horizontal axis is $F\cos\theta$.  A spectrum obtained at a field angle of $\theta$ degrees is shifted vertically by $\theta$ (in the units of the vertical axis).  The amplitudes of some spectra are magnified ($\times 5$) or reduced ($\times 0.5$) for the clarity as indicated in the figure.  The main spectra showing the frequency range 0 $< F\cos\theta <$ 5 kT are Fourier transforms over a field window 7 $< B < $ 17.65 T, while the superimposed spectra for the range 2.9 $ \lesssim F\cos\theta \lesssim$ 4.5 kT are transforms over a narrower field window on the high field side, 10 $< B < $ 17.65 T, so that the $\zeta$ peaks are enhanced.  The dotted lines drawn for the $\alpha$ and $\epsilon$ frequencies are based on Yamaji model, and the shading in (c) indicates a frequency spread due to a $\pm$1\degree~distribution of the $c$ axis orientation (see text).  The lines drawn for the $\zeta$ frequency are a guide to the eyes.  No fundamental frequencies were observed above $F\cos\theta$ = 5 kT.}   
\end{figure*}

\begin{figure}
\begin{center}
\includegraphics[width=7cm]{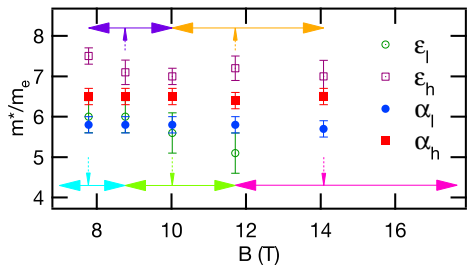}
\end{center}
\caption{\label{fig4}(color online).  Effective masses versus magnetic field ($B \parallel c$).  Field ranges used for data points are indicated by horizontal bars with arrows at both ends.}   
\end{figure}

The electronic band structure of KFe$_2$As$_2$ was calculated within the local density approximation (LDA) by using a full potential LAPW (FLAPW) method.  We used the program codes TSPACE\cite{Yanase1995} and KANSAI-06.  The experimental crystal structure \cite{Rozsa81ZNB} including the atomic position $z_{As}$ of As was used for the calculation.  Figure~\ref{fig1} shows the calculated band structure, density of states (DOS), and FS.  The calculated dHvA frequencies $F$ and band masses $m_{band}$ for $B \parallel c$ are tabulated in Table~\ref{Tab1}.  The DOS at the Fermi level ($E_F$) is 58.4 states/(Ry f.u.), which corresponds to the Sommerfeld coefficient of $\gamma_{band}$ =  10.1 mJ/K$^2$mol-f.u.  As can be seen from the DOS plot, the states near $E_F$ are mostly derived from the Fe 3$d$ orbitals.  Four bands 31--34 cross $E_F$, and the FS consists of three concentric hole cylinders at the $\Gamma$ point of the BZ (bands-32, 33, and 34), small hole cylinders near the zone boundary (band-34), and a small hole pocket at Z (band-31).  Although this FS is rather different from a previously calculated one \cite{Singh09PRB}, which has an electron cylinder at X instead of the hole cylinders around X, it is consistent with that determined by ARPES except that the small pocket at Z is not experimentally observed.

Single crystals of KFe$_2$As$_2$ were grown from a potassium flux.  Resistivity measurements on some of the grown crystals show the residual resistivity ratios of $\sim$600.  The dHvA measurements were performed in a dilution refrigerator and superconducting magnet by using the field modulation technique \cite{Shoenberg84}.  The modulation frequency and amplitude are 67.1 Hz and 10.4 mT, respectively.  The field direction measured from the $c$ axis is denoted by $\theta$, and if necessary a subscript is attached to indicate the rotation plane.

Figure~\ref{fig2} shows the Fourier transform of dHvA oscillations at $\theta_{(010)}$ = 36.7\degree.  Six fundamental frequencies $\epsilon_{l, h}$, $\alpha_{l, h}$, and $\zeta_{l, h}$ are resolved.  Figure~\ref{fig3} shows the field-direction dependence of the spectra.  Note that the horizontal axis is $F\cos\theta$.  Effective masses of electrons were determined from the temperature dependence of dHvA oscillation amplitudes between 0.1 and 0.43 K (0.2 K for $\zeta_h$) for $B \parallel c$ (Table~\ref{Tab1}).  We also examined the field dependence of the effective masses for the $\epsilon$ and $\alpha$ frequencies; the masses are constant within the errors in the investigated field range as shown in Fig.~\ref{fig4}.  This is in stark contrast to a recent report of strong field dependence of the coefficient $A$ of the resistivity.\cite{Dong10PRL}

In the case of a purely two-dimensional (2D) electronic structure, the FS would be a straight cylinder showing a single dHvA frequency $F$, and $F\cos\theta$ would be constant irrespective of the field direction.  In reality, there is a $c$-axis dispersion of the electronic band energy, which leads to corrugation of the cylindrical FS and produces at least two dHvA frequencies corresponding to the maximal and minimal FS cross sections.  The simplest case where the $c$-axis dispersion takes a form of $\cos I_c k_z$ ($I_c$ is the interlayer distance) was considered by Yamaji \cite{Yamaji89JPSJ}, and the angle dependence of the dHvA frequencies were derived.

We assign each of the three pairs of the observed frequencies $\epsilon_{l, h}$, $\alpha_{l, h}$, and $\zeta_{l, h}$ to a corrugated cylindrical FS.  The angle dependence of $\epsilon_{l, h}$ is consistent with Yamaji model as shown by the dotted lines in Figs.~\ref{fig3}(a) and (b).  The angle dependence of $\alpha_{l, h}$ can also be explained by Yamaji model if distribution of the $c$-axis orientation in the sample is considered.  The shading in Fig.~\ref{fig3}(c) indicates the frequency spread calculated for the $\pm$1\degree~distribution of the $c$-axis orientation.  As the field is tilted from the $c$ axis, the two frequencies $\alpha_{l, h}$ show a tendency to approach each other, but above around 35\degree~they split into small peaks, which spread out as expected from the calculated frequency spread.  Despite the frequency spread, we notice an important characteristic of Yamaji model appearing at high angles; namely, the frequency peaks are enhanced near $\theta$ = $\pm$52.5\degree.  This is because at this magic angle the maximum and minimum frequencies are expected to coincide and the whole region of the FS contributes to the single-frequency dHvA oscillation, resulting in an enhanced oscillation amplitude.  The $\zeta$ frequencies exhibit a distinctive angle dependence.   The minimum frequency $\zeta_l$ shows much less variation of $F\cos\theta$ than the maximum one $\zeta_h$.  This indicates that the corresponding FS cylinder is close to a straight one near the minimal cross section, and that it bulges out rather locally around the maximal cross section.  The effective masses of $\zeta_l$ and $\zeta_h$ differ considerably for $B \parallel c$ (Table~\ref{Tab1}), but they become close as the field is tilted; $m^*$ = 11.1(4) and 12.5(7) $m_e$, respectively, at $\theta_{(010)}$ = 36.7\degree.  This also indicates that the modification of the FS is local.  A similar but more extreme bulge has been observed for the larger hole FS of SrFe$_2$P$_2$ \cite{Analytis09PRL}.

In addition to the above frequency branches, many frequencies appear between $\alpha_l$ and $\zeta_l$ for field directions near $B \parallel c$ (Fig.~\ref{fig3}).  These frequencies can be expressed as $F_{\alpha_l} + n \Delta F$ ($n$ = 3, 4, 5, \dots), where $\Delta F$ = $(F_{\zeta_l}-F_{\alpha_l})/12$.  The frequencies $n$ = 1 and 2 are not clearly resolved probably because they are overlapped by the $\alpha_h$ frequency.  Frequencies expressed as $2F_{\alpha_l} + n \Delta F$ ($n$ = 1, 2, 3, \dots) are also observed (small peaks appearing for $F\cos\theta \gtrsim$ 4.6 kT in Fig.~\ref{fig3}).  The effective masses of the frequencies $F_{\alpha_l} + n \Delta F$ range between those of $\alpha_l$ and $\zeta_l$, and, as the field is decreased, their amplitudes are damped considerably faster than those of $\alpha_l$ and $\zeta_l$.  Therefore, they are most likely combination frequencies involving magnetic breakdowns.  If the $\alpha$ and $\zeta$ FS's are close enough for magnetic breakdown to occur at some point in the BZ, there will be four such points because of the tetragonal symmetry.  This can explain the frequencies with $n$ = 3, 6, and 9, and these frequencies are stronger than their respective neighboring frequencies.  The origin of the observed three-times finer splitting however remains unclear. 

We identify our $\epsilon$, $\alpha$, and $\zeta$ cylinders with the $\epsilon$ FS and the inner and the outer FS of the quasi-degenerate $\alpha$ FS observed by the ARPES measurements \cite{Sato09PRL, Yoshida09PC}, respectively, and also with the band-34 small hole cylinder near X, the bands-32 and 33 hole cylinders at $\Gamma$ predicted by the band structure calculation, respectively.  The volumes of the observed cylinders are estimated from the averages of the minimum and maximum frequencies for $B\parallel c$ to be 1.1, 8.4, and 13\% of the BZ, respectively.  The total observed volume is (4 $\times$ 1.1 + 8.4 + 13 =) 26\%.  Since KFe$_2$As$_2$ is uncompensated, the total volume has to be 50\%.  We attribute the unobserved 24\% to the $\beta$ FS.\cite{beta_or_3alpha}  Using a 2D approximation, we can estimate the specific heat $\gamma$ from the measured effective masses.  The observed FS's contribute 67mJ/K$^2$mol-f.u to $\gamma$ (with a roughly estimated error of $\pm$10\%).  This is to be compared with the measured single-crystal value of 93 mJ/K$^2$mol-f.u. \cite{Fukazawa09PC}, and the difference is ascribed to the $\beta$ cylinder.

Very recently, angle-dependent magnetoresistance oscillations (AMRO's) have been observed in KFe$_2$As$_2$ \cite{Kimata10PC}.  Two FS's are resolved and the estimated FS sizes are 12 and 17\% of the BZ with a relative error of $\pm20$\%.  They can be identified with the present $\alpha$ and $\zeta$ FS's.

There are large discrepancies between the measured and calculated dHvA frequencies (Table~\ref{Tab1}).  As for the $c$-axis dispersion, those of the $\alpha$ and $\zeta$ bands are considerably smaller and larger than calculated, respectively.  In order to grasp how accurate band structure calculations usually are, we refer to our previous dHvA measurements on BaNi$_2$P$_2$ with the same ThCr$_2$Si$_2$ structure \cite{Terashima09JPSJ_BaNi2P2}.  The band structure calculation for BaNi$_2$P$_2$ was done with the same procedure and codes as the present one.  Discrepancies between the measured and calculated dHvA frequencies for $F > 0.5$ kT were 7\% at most and were typically much smaller.  The remarkable discrepancies found in KFe$_2$As$_2$ suggest the necessity of treating electronic correlations beyond the LDA as suggested by previous theoretical studies.  Ref.~\citen{Ortenzi09PRL} pointed out that self-energy effects arising from the spin-mediated interband interaction (mainly between electrons and holes) could cause band shifts relative to LDA calculations, while Ref.~\citen{Aichhorn09PRB} showed that when correlations were treated by the dynamical mean field theory the crystal field splitting of the Fe 3$d$ orbitals were modified, which would alter the FS.  Since there is no electron FS and SF's are suppressed \cite{Fukazawa09JPSJ_KFA}, the latter seems more relevant to KFe$_2$As$_2$.

The observed mass enhancements are about 3 to 7 for the $\alpha$ and $\zeta$ FS's.\cite{epsilon_mass}  They are broadly consistent with the specific-heat mass enhancement $\gamma_{exp}/\gamma_{band}$ = 9.2.  These enhancements are comparable to or even larger than the value of $\sim$4 found in the $T_c$ $\sim$ 25 K compound BaFe$_2$(As$_{0.59}$P$_{0.41}$)$_2$ \cite{Shishido09condmat}.  Considering the suppressed SF's and the field independence of the effective masses (Fig.~\ref{fig4}), we do \textit{not} ascribe the main origin of the mass enhancements specifically to low-energy SF's, but we ascribe it to band narrowing due to more general electronic correlations arising from the local Coulomb interaction on the Fe 3$d$ shell.  The importance of such high-energy correlations has been pointed out from recent optical studies on the iron pnictides \cite{Qazilbash09nphys}.  It is also interesting to note a recent theoretical work,\cite{Ikeda10condmat} which suggests that the mass enhancement increases with hole doping in iron pnictides.

In summary, we have observed dHvA oscillations from the $\epsilon$ hole cylinders near the X point of the BZ and the $\alpha$ and $\zeta$ hole cylinders at $\Gamma$.  Effective masses of electrons range from 6 to 18$m_e$ for $B \parallel c$, and the mass enhancements are 3 to 7 for the $\alpha$ and $\zeta$ cylinders.  The specific-heat mass enhancement is 9.2.  The unusual discrepancies between the observed and calculated FS areas and the large mass enhancements clearly demonstrate the necessity of including electronic correlations beyond the LDA in understanding the electronic structures of the iron pnictides superconductors.


\begin{thebibliography}{10}
\bibitem{beta_or_3alpha}
Since the $\beta$ frequency is expected to appear near the third harmonic of $\alpha$ ($3 \times 8.4$ = 25.2\%), the assignment of 3$\alpha$ has to be done with caution.  In the present study, we have verified at two field directions ($B \parallel c$ and $\theta_{(010)}$ = 36.7\degree) that the effective masses associated with 3$\alpha_h$ are three times larger than those of $\alpha_h$ as is expected for the third harmonic.
\bibitem{epsilon_mass}
The mass enhancements for the $\epsilon$ FS are probably overestimated because, if the band is shifted up to match the measured and calculated FS sizes, a band minimum along the $\Delta$ line will approach $E_F$ and $m_{band}$ will be increased.
\end{thebibliography}
\end{document}